\documentclass[11pt]{article}
\usepackage{graphics,amsmath,amssymb,amsfonts}
\usepackage{graphics}

\newtheorem{Lemma}{Lemma}[section]
\newtheorem{Theorem}{Theorem}[section]
\newtheorem{Proposition}{Proposition}[section]
\numberwithin{equation}{section}
\def \Tr#1{{\rm Tr}(#1)}
\title{Wronskian solutions to the KdV equation via B\"acklund transformation}
\author{Qi-fei Xuan\footnote{E-mail:
xuanzi001@yahoo.com.cn },  ~~ Mei-ying Ou, ~~ Da-jun Zhang\footnote{Corresponding author. E-mail:
djzhang@staff.shu.edu.cn }
\\{\small \it Department of Mathematics, Shanghai University,
Shanghai 200444, P.R. China}}
\begin{document}
\maketitle

\begin{abstract}
In the paper we discuss the
B\"acklund transformation of the KdV equation
between solitons and solitons, between negatons and negatons,
between positons and positons, between rational solution and rational solution,
and between complexitons and complexitons.
We investigate the conditions that Wronskian entries satisfy for the
bilinear B\"acklund transformation of the KdV equation.
By choosing suitable Wronskian entries and the   parameter in the bilinear B\"acklund transformation,
we obtain transformations between many kinds of solutions.
\end{abstract}

\vskip 18pt
\section{Introduction}

The Wronskian can be considered as a bridge connecting with
many classical methods in soliton theory. This is not only because soliton solutions in
Wronskian form can be obtained from the Darboux transformation\cite{Matveev-book},
Sato theory\cite{Sato,Sato-elementary} and Wronskian technique\cite{Freeman-Nimmo-KP}-\cite{Freeman-IMA},
but also because  the exponential polynomial for $N$-solitons derived from
Hirota method\cite{Hirota-1971,Hirota-book}
and the matrix form given by the Inverse Scattering Transform\cite{GGKM,Ablowitz-book-1981} can be
transformed into a Wronskian by extracting some exponential factors.
The special structure of a Wronskian contributes simple
forms of its derivatives, and this admits solution verification by
direct substituting Wronskians into a bilinear soliton equation
or a bilinear B\"acklund transformation(BT). This approach is referred to as
Wronskian technique\cite{Freeman-Nimmo-KP}.
In the approach a bilinear soliton equation is some algebraic identity
provided that Wronskian entry vector satisfies some differential equation set
which we call Wronskian condition.
A useful generalization for the technique was made by
Sirianunpiboon, Howard and Roy (SHR)\cite{Sirianunpiboon-1988}
who substituted a triangular coefficient matrix
for the original diagonal one in the Wronskian condition for the KdV equation.
This generalization enables many solution such as rational solutions,
positons\cite{Matveev-positon-kdv1,Matveev-positon-kdv2},
negatons\cite{negaton-positon-96}, complexitons\cite{Ma-complexiton,breather-KdV} and mixed solutions
to be expressed in Wronskian form.
In two recent papers\cite{Ma-You-KdV,Zhang-Wronskian}, solutions to the KdV equation
were reviewed in term of Wronskian on the basis of SHR's generalization.
In \cite{Ma-You-KdV} Ma and You first investigated all kinds of Wronskian entry vectors related to
different canonical forms of the coefficient matrix in the Wronskian condition.
In \cite{Zhang-Wronskian} Zhang formulated the Wronskian technique as four steps, (i.e.,
finding the most possible wide Wronskian conditions, getting all possible Wronskian entry vectors,
describing relations between different kinds of solutions,
and discussing  parameter effects and dynamics of the solutions obtained),
and gave detailed discussions for the first three steps.

Both SHR's generalization\cite{Sirianunpiboon-1988} and the two recent
reviews\cite{Ma-You-KdV,Zhang-Wronskian} are based on bilinear equations but not BTs.
It is well known that besides bilinear equations, the Wronskian verifications
can also be achieved through bilinear BTs.
Bilinear BT was first found by Hirota in 1974\cite{Hirota-BT-1974}.
In general, there is a parameter in a bilinear BT,
and by putting special value to the parameter
the BT can provide a transformation between $(N-1)$-solitons and $N$-solitons.
Two natural questions are whether SHR's generalization can be achieved via
bilinear BTs and how to choose the parameter in a BT so that we get
transformations between positons, negatons, rational solutions and complexitons.
So far for those solutions other than solitons which are derived via bilinear BTs,
only some nonlinear superaddition formulae for rational solutions were
reported\cite{Hu-Rational-1995}.

In this paper we  aim to investigate Wronskian solutions to the KdV equation via
its bilinear BT. Our work will mainly focus on what the Wronskian condition is under the
bilinear BT  and how the parameter in the BT is related to
positons, negatons, rational solutions and complexitons.
We will show that the BT can
provide transformations between solitons and solitons, between negatons and negatons,
between positons and positons, between rational solution and rational solution,
and between complexitons and complexitons.

In the paper we will first in Sec.2 investigate few properties
of a kind of matrix with special form.
Then in Sec.3 we discuss Wronskian solutions to the KdV equation
via bilinear BT.
Finally in Sec.4 we give Wronskian entries and transformations for many kinds of solutions.


\section{A kind of matrix with special form and their properties}

We consider the following complex $s\times s$ matrix set
\begin{equation}
\mathcal{G}_{s}=\biggl\{A_{s\times s}~\Bigl|
~A_{s\times s}=\Biggl(
 \begin{array}{cc}
                \hat{A} & \bf{0} \\
                \alpha & a_{ss}\\
               \end{array}
              \Biggr)\biggr\},
\label{A_s}
\end{equation}
where $\hat{A}$ is an $(s-1)\times(s-1)$ matrix with arbitrary complex numbers as entries,
the complex vector $\alpha=(\alpha_{1},\cdots,\alpha_{s-1})$,
$\bf{0}$ stands for an $s-1$ order column zero vector,
and $a_{ss}$ is a complex scalar.

Noting that $A_{s\times s}$ is a block lower triangular matrix,
we immediately reach to the following property.

\begin{Proposition}
\label{Pro-1}
$\mathcal{G}_{s}$ defined by \eqref{A_s} forms a semigroup with identity
with respect to matrix multiplication and inverse.
\end{Proposition}

Since any square matrix is similar to a lower triangular one
on the complex number filed $\mathbb{C}$, we have
\begin{Proposition}
 \label{Pro-3}
 For any given $A\in \mathcal{G}_{s}$, there exists a transformation matrix $B\in \mathcal{G}_{s}$
 such that
\begin{equation*}
A=B^{-1}CB,
\end{equation*}
where $C$ is an $s\times s$ lower triangular matrix.
\end {Proposition}

\noindent{\it Proof:}~ Suppose that $\hat{B}$ is a non-singular $(s-1)\times (s-1)$
transformation matrix such that $\hat{A}$ is similar to a lower triangular matrix $\hat{C}$,
i.e., $\hat{A}=\hat{B}^{-1} \hat{C} \hat{B}$. Then for $A$ we can take
$B=\Bigl(\begin{smallmatrix}\hat{B} & \bf{0} \\\bf{0} & 1
\end{smallmatrix}\Bigr)$
and consequently
$C=B A {B}^{-1} =\Bigl(\begin{smallmatrix}\hat{C} & \bf{0} \\\alpha \hat{B}^{-1} & 1
\end{smallmatrix}\Bigr)$ which is an $s\times s$ lower triangular matrix.
\hfill $\Box$

\begin{Proposition}
\label{Pro-4}
 For any given $A\in \mathcal{G}_{s}$, there exists a matrix $T
\in  \mathcal{G}_{s}$ satisfying
$$T^{-1}AT=\Gamma,$$
where $\Gamma$
is defined as the following form
\begin{equation}
\Gamma=\left(
           \begin{array}{cc}
             \Gamma_{s-1}& 0 \\
             \gamma & a_{ss} \\
           \end{array}
         \right)
\label{quasi-cano}
\end{equation}
with $\Gamma_{s-1}$ being the canonical form of $ \hat{A}$ and
$\gamma$ being an $(s-1)$-order row vector. Here we call
\eqref{quasi-cano} quasi-canonical form of $A$.
\end{Proposition}

\noindent{\bf Proof: }For arbitrary  $(s-1)\times(s-1)$ matrix $
\hat{A}$, there exists a nonsingular $(s-1)\times(s-1)$ matrix $
\hat{T}$ satisfying $ \hat{T}^{-1} \hat{A} \hat{T}=\Gamma_{s-1}$.
Let
$$T=\left(
 \begin{array}{cc}
   \hat{T} & 0\\
   0 & 1 \\
   \end{array}
   \right),$$
then we have $$T^{-1}AT=\left(
 \begin{array}{cc}
  \hat{T}^{-1} & 0\\
   0 & 1 \\
   \end{array}
   \right)
\left(
 \begin{array}{cc}
                 \hat{A} & \bf{0} \\
                \bf{\alpha} & a_{ss}\\
               \end{array}
              \right)
\left(
 \begin{array}{cc}
   \hat{T} & 0\\
   0 & 1 \\
   \end{array}
   \right)
=\left(
   \begin{array}{cc}
     \Gamma_{s-1} & 0 \\
     \alpha  \hat{T}& a_{ss} \\
   \end{array}
 \right)=\Gamma,$$
 where
$\alpha
 \hat{T}=\gamma=(\gamma_1,\gamma_2,\cdots,\gamma_{s-1})$.
 \hfill $\Box$

Now we expand $\mathcal{G}_s$ to a function semi-group
$\mathcal{G}_s[t]$ which consists of $A_{s\times s}$
defined as \eqref{A_s} but each non-zero entry  of $A_{s\times s}$
is a functions of $t$ instead.
For the function matrix in $\mathcal{G}_s[t]$, we have the following result.

\begin{Proposition}
\label{Prop-2} Suppose that
$B(t)=(b_{ij}(t))\in \mathcal{G}_N[t]$ with $b_{NN}(t)\equiv 0$ and each $b_{ij}(t)\in C[a,b]$
where $a$ and $b$ can be infinite. Then, there exists a
non-singular $N \times N$ $t$-dependent matrix $H(t)=(h_{ij}(t))\in \mathcal{G}_N[t]$, satisfying
\begin{equation}
H(t)_{t}=-H(t)B(t).
\label{H=BH}
\end{equation}
\end{Proposition}

\noindent{ \bf{Proof: }}The proof is similar to the one for Proposition 3.6 in Ref.\cite{Zhang-Wronskian}.
We consider the following homogeneous linear
ordinary differential equations
\begin{equation}
h_t(t)=-B^T(t)h(t),
\label{h=Bh}
\end{equation}
where $h(t)$ is an $N$-order column vector function of $t$. Then, for any given constant
number set
$(t_0,\tilde{h}_{j1},\tilde{h}_{j2},\cdots,\tilde{h}_{jN})$, under
the condition of the proposition,
 \eqref{h=Bh} has unique solution vector
\begin{equation*}
h_j(t)=(h_{j1}(t),h_{j2}(t),\cdots,h_{jN}(t))^T
\end{equation*}
satisfying
$h_j(t_0)=(\tilde{h}_{j1},\tilde{h}_{j2},\cdots,\tilde{h}_{jN})^T$
for each $j=1,2,\cdots N$.
Noting that $B(t)\in \mathcal{G}_N[t]$ and $b_{NN}(t)\equiv 0$,
\eqref{h=Bh} suggests
\begin{equation*}
[h_{jN}(t)]_{t}=0,~~ j=1,2,\cdots,N.
\end{equation*}
Now we specially take $(\tilde{h}_{js})_{N \times N}$ which belongs
to $\mathcal{G}_N$ and is nonsingular, and meanwhile, we take
$h_{jN}(t)=0$ for $j=1,2,\cdots,N-1$ and
$h_{NN}(t)=\tilde{h}_{NN}\neq 0$. That guarantees us to get a
nonsingular function matrix
\begin{equation*}
H(t)=(h_{1}(t),h_{2}(t),\cdots,h_{N}(t))^T
\end{equation*}
which solves \eqref{H=BH}. Thus we complete the proof.
\hfill $\Box$

\section{Wronskian solutions to the KdV equation via bilinear BT}

\subsection{ Bilinear BT of the KdV equation }

The KdV equation is
  \begin{equation}
  u_{t}+6uu_{x}+u_{xxx}=0
\label{KdV}
  \end{equation}
with bilinear form\cite{Hirota-1971}
\begin{equation}
(D_{t}D_{x}+D_{x}^4)f \cdot f=0
\label{KdV-bilinear}
\end{equation}
under the transformation
\begin{equation}
u= 2 (\ln f)_{xx}, \label{trans-f}
\end{equation}
where $D$ is the well-known Hirota's bilinear operator defined by\cite{Hirota-book}
\begin{displaymath}
 D_{t}^mD_{x}^na(t,x)\cdot b(t,x) =\frac{\partial^m}{{\partial s}^m}\frac{\partial^n}{{\partial y}^n}
a(t+s,x+y)b(t-s,x-y)|_{s=0,y=0},~~ m,n=1,2,\cdots .
\end{displaymath}
From \eqref{KdV-bilinear}, Hirota gave the bilinear BT of the KdV equation\cite{Hirota-BT-1974} as
\begin{subequations}
\label{KdV-BT}
\begin{eqnarray}
         D_{x}^2f \cdot g- h f \cdot g =0, \label{KdV-BT-1}\\
        (D_{x}^3+D_{t}+3h D_{x})f \cdot g=0,
        \label{KdV-BT-2}
  \end{eqnarray}
   \end{subequations}
where $h$ is a parameter.

\subsection{ Wronskian solutions  to the KdV equation via bilinear BT }

An ${N}\times{N}$ Wronskian is defined as
\begin{displaymath}
 W(\phi_1, \phi_2, \cdots, \phi_N)
   = \left|\begin{array}{cccc}
   \phi_1 &\phi_1^{(1)} &\cdots &\phi_{1}^{(N-1)}\\
   \cdots &\cdots &\cdots &\cdots\\
    \phi_N &\phi_N^{(1)} &\cdots &\phi_N^{(N-1)}
    \end{array}\right|,
    \end{displaymath}
 where $\phi_j^{(l)}=\partial^{(l)}\phi_{j}/{\partial x}^l$.
It can be expressed by the following compact form
\begin{displaymath}
W(\phi)=|\phi, \phi^{(1)}, \cdots, \phi^{(N-1)}|=|0, 1, \cdots,
N-1|=|\widehat{N-1}|,
\end{displaymath}
where $\phi=(\phi_1, \phi_2, \cdots, \phi_N)^T$.

If $f=|\widehat{N-1}|$ and each entry $\phi_j$
satisfies\cite{Freeman-Nimmo-KP}
\begin{equation}
\begin{array}{r}
-\phi_{j,xx}=\lambda_j \phi_j, \\
\phi_{j,t}=-4\phi_{j,xxx},
\end{array}
\label{Wrons-cond-1}
\end{equation}
with arbitrary parameter $k_j$,
then such an $f$ solves the bilinear KdV equation \eqref{KdV-bilinear}.
Besides, under the same Wronskian condition as \eqref{Wrons-cond-1}, the
bilinear BT \eqref{KdV-BT} provides a transformation\cite{Freeman-Nimmo-KP} between
$g=|\widehat{N-1}|$ and $f=|\widehat{N-2},\tau_{N}|$ where we take
$h=\lambda_N$ and
\begin{equation}
\tau_{j}=(\delta_{j, 1}, \delta_{j, 2}, \cdots, \delta_{j,
N}), ~~~ \delta_{j, i}=\biggl\{\begin{array}{c}
             0, ~~ i\neq j, \\
             1, ~~ i=j.
             \end{array}
\label{tau}
\end{equation}
In this case, $g$ and $f$ provide $N$-solitons and $(N-1)$-solitons of the KdV equation \eqref{KdV}
through
\begin{equation}
u= 2(\ln g)_{xx} \label{trans-g}
\end{equation}
and \eqref{trans-f}, respectively.

In a recent review\cite{Zhang-Wronskian} it is shown that
the bilinear KdV equation \eqref{KdV-bilinear} admits
Wronskian solution
\begin{equation}
f=W(\phi)=|\widehat{N-1}|
\end{equation}
and the Wronskian condition is
\begin{equation}
\begin{array}{r}
 -\phi_{xx}=A(t)\phi, \\
  \phi_{t}=-4\phi_{xxx}+B(t)\phi,
\end{array}
\end{equation}
where $A(t)$ and $B(t)$ are two arbitrary $N\times N$ matrices of $t$ but
independent of $x$ and satisfy
\begin{equation}
A(t)_{t}+[A(t), B(t)]=0,
\end{equation}
in which $[A(t), B(t)]=A(t)B(t)-B(t)A(t)$.

In what follows we investigate the Wronskian condition on the basis of
the bilinear BT \eqref{KdV-BT}. Let us first give the following theorem.

\begin{Theorem}
\label{Theor-1}
~Wronskian solutions to the bilinear BT \eqref{KdV-BT}
can be given by
\begin{equation}
g=|\widehat{N-1}|,~~ f=|\widehat{N-2},\tau_{N}|, \label{f-g}
\end{equation}
where $\tau_{N}$ is defined as \eqref{tau}
and the Wronskian condition is
\begin{subequations}
\begin{eqnarray}
 -\phi_{xx}=A(t)\phi,
 \label{Wr-Cd-a}\\
  \phi_{t}=-4\phi_{xxx}+B(t)\phi,
 \label{Wr-Cd-b}
\end{eqnarray}
\label{KdV-BT-condition}
\end{subequations}
$A(t)=(a_{ij}(t))_{N\times N}\in \mathcal{G}_{N}[t]$, $B(t)=(b_{ij}(t))_{N\times N}\in \mathcal{G}_{N}[t]$
with $b_{NN}(t)\equiv 0$, satisfying
\begin{equation}
A(t)_{t}+[A(t), B(t)]=0,
\label{KdV-AB}
\end{equation}
and in \eqref{KdV-BT} we take  $h=-a_{NN}(t)\equiv -a_{NN}$ which is
a constant due to \eqref{KdV-AB}.\end{Theorem}

%
%

Before giving the proof we consider the following Lemmas.

\begin{Lemma}
{\rm\cite{Freeman-Nimmo-KP}}
\label{Lemma-1}
~Suppose that $M$ is an $N\times (N-2)$ matrix, and $a, b, c, d$ are $N$-order
column vectors. Then, we have
\begin{equation*}
|M, a, b||M, c, d|-|M, a, c||M, b, d|+|M, a, d||M, b, c|=0.
\end{equation*}
\end{Lemma}

\begin{Lemma}
{\rm \cite{Zhang-Wronskian,Zhang-Hietarinta}}
\label{Lemma-2}
~ Suppose that $\Xi$ is an
 $N\times N$ matrix with column vector set $\{\Xi_j\}$; $\Omega$ is an
 $N\times N$ operator matrix with column vector set $\{\Omega_j\}$ and
 each entry $\Omega_{js}$ is an operator. Then
 we have
\begin{equation}
\sum^N_{j=1} |\Omega_j * \Xi| =\sum^N_{j=1}|(\Omega^T)_{j} *
\Xi^T|,
\end{equation}
 where for any $N$-order column vectors $A_j$ and $B_j$ we define
\begin{equation*}
A_j \circ B_j=(A_{1 j}B_{1 j}, ~A_{2 j}B_{2 j}, \cdots, A_{N
j}B_{N j})^T
\end{equation*}
 and
\begin{equation*}
|A_j * \Xi|=|\Xi_1, \cdots, \Xi_{j-1}, ~A_j \circ\Xi_j, ~\Xi_{j+1},
\cdots, \Xi_{N}|
\end{equation*}
\end{Lemma}

By virtue of Lemma \ref{Lemma-2} we can have several identities. For example,
taking $\Xi=|\widehat{N-1}|$ and $\Omega_{js}=\partial_{x}^2$ in Lemma \ref{Lemma-2}
 and under the Wronskian condition \eqref{Wr-Cd-a},  we
 can get
\begin{equation}
-\Tr{A(t)}|\widehat{N-1}|=|\widehat{N-2}, N+1|-|\widehat{N-3},
N-1, N|, \label{(+11)}
\end{equation}
where $\Tr{A(t)}$ is the trace of $A(t)$. Similarly , let
$\Xi=|\widehat{N-2}, \tau_{N}|$ and $\Omega_{js}=\partial_{x}^2$,
 we have
\begin{equation}
-\Tr{A(t)}|\widehat{N-2}, \tau_{N}|=|\widehat{N-3}, N,
\tau_{N}|-|\widehat{N-4}, N-2, N-1, \tau_{N}|
-\sum_{j=1}^Na_{jN}|\widehat{N-2}, \tau_{j}|; \label{(+21)}
\end{equation}
let
 $\Xi=|\widehat{N-2},N|$ and $\Omega_{js}=\partial_{x}^2$,
we have
\begin{equation}
-\Tr{A(t)}|\widehat{N-2}, N|=|\widehat{N-2}, N+2|-|\widehat{N-4},
N-2, N-1, N|; \label{(+31)}
\end{equation}
let $\Xi=|\widehat{N-3},N-1,\tau_{N}|$ and
$\Omega_{js}=\partial_{x}^2$, we have
\begin{equation}
  \begin{array}{lll}
 -\Tr{A(t)}|\widehat{N-3}, N-1, \tau_{N}| &=& |\widehat{N-3}, N+1, \tau_{N}|-|\widehat{N-5}, N-3, N-2, N-1, \tau_{N}| \\
  ~ & ~& -\displaystyle \sum_{j=1}^Na_{jN}|\widehat{N-3}, N-1, \tau_{j}|.\\
 \end{array} \label{(+41)}
  \end{equation}
These equalities will play  important roles in our proof.

\begin{Lemma}
\label{Lemma-3}
~ Consider $N \times N$ matrices
$$g(\phi)=|\widehat{N-1}|, ~~f(\phi)=|\widehat{N-2}, \tau_{N}|, $$
where the entry vector $\phi$ satisfies
$$\phi_{t}=B(t)\phi,$$
and $B(t)=(b_{ij}(t))_{N\times N}$ is an $N \times N$ $t$-dependent matrix. Then,
we have
\begin{equation}
 g_{t}=\Tr{B(t)}g, ~~
 f_{t}=\Tr{B(t)}f-\sum_{j=1}^N b_{jN}|\widehat{N-2}, \tau_{j}|.
 \end{equation}
In addition, if taking $b_{jN}(t)=0$ for $j=1, 2 \cdots, N$, then $D_{t}g\cdot f=0$.
\end{Lemma}

Now, we give the proof of Theorem \ref{Theor-1}.

\noindent{\bf Proof for Theorem \ref{Theor-1}:}
Instead of \eqref{KdV-BT-condition} we first consider the following
Wronskian condition,
\begin{equation*}
\begin{array}{r}
 -\phi_{xx}=A(t)\phi, \\
  \phi_{t}=-4\phi_{xxx},
\label{Wrons-cond-2}
\end{array}
\end{equation*}
where $A(t)=(a_{ij}(t))_{N \times N}$ is an arbitrary $t$-dependent
 $N \times N$ matrix.

Substituting \eqref{f-g} into   \eqref{KdV-BT-1} yields
\begin{equation}
\begin{array}{rl}
 ~& g_{xx}f-2g_{x}f_{x}+gf_{xx}-hgf \\
 =& |\widehat{N-3}, N-1, N||\widehat{N-2}, \tau_{N}|
 +|\widehat{N-2}, N+1||\widehat{N-2},\tau_{N}|
 -2|\widehat{N-2}, N||\widehat{N-3}, N-1, \tau_{N}|\\
  ~ &+ |\widehat{N-1}||\widehat{N-4}, N-2, N-1, \tau_{N}|
  +|\widehat{N-1}||\widehat{N-3}, N, \tau_{N}|-h|\widehat{N-1}||\widehat{N-2}, \tau_{N}|.
  \end{array}\label{(8888)}
  \end{equation}
By means of (\ref{(+11)}), (\ref{(+21)}) and the identity
$$[\Tr{A(t)}|\widehat{N-1}|]|\widehat{N-2}, \tau_{N}|-|\widehat{N-1}|[\Tr{A(t)}|\widehat{N-2}, \tau_{N}|]=0, $$
(\ref{(8888)}) can be simplified to
$$\begin{array}{rl}
~& g_{xx}f-2g_{x}f_{x}+gf_{xx}-hgf \\
=& 2|\widehat{N-3}, N-1, N||\widehat{N-2}, \tau_{N}|
 -2|\widehat{N-2}, N||\widehat{N-3}, N-1, \tau_{N}|\\
 ~& +2|\widehat{N-1}||\widehat{N-3}, N, \tau_{N}|
 -|\widehat{N-1}|\widehat{N-2}, \hat{\tau}_{N}| ,
\end{array}$$
where
\begin{equation}
\hat{\tau}_{N}=(a_{N1}(t), ~ \cdots,~ a_{N(N-1)}(t),~
a_{NN}(t)+h)^T.
\end{equation}
Then, in light of Lemma \ref{Lemma-1} we have the identity
\begin{equation}
|\widehat{N-3}, N-1, N||\widehat{N-2}, \tau_{N}|
 -|\widehat{N-2}, N||\widehat{N-3}, N-1, \tau_{N}|
 +|\widehat{N-1}||\widehat{N-3}, N, \tau_{N}|=0,
 \end{equation}
and by this identity  \eqref{(8888)} is  reduced to
$$g_{xx}f-2g_{x}f_{x}+gf_{xx}-hgf=|\widehat{N-1}||\widehat{N-2}, \hat{\tau}_{N}|.$$
Obviously, to make the above formula zero, we need to take
$\hat{\tau}_N=0$, i.e., $A(t) \in \mathcal{G}_{N}$ and
$h=-a_{NN}(t)$. Thus \eqref{KdV-BT-1} has been proved.

For \eqref{KdV-BT-2}, making use of (\ref{(+11)}), (\ref{(+41)}) and the identity
\begin{equation*}
\Bigl (\Tr{A(t)}|\widehat{N-1}|\Bigr )|\widehat{N-2}, \tau_{N}|_{x}
-|\widehat{N-1}|\left (\Tr{A(t)}|\widehat{N-2},\tau_{N}|_{x}\right )=0,
\end{equation*}
and noting $\hat{\tau}_{N}=0$, we have
\begin{equation*}
\begin{array}{rl}
l.h.s. \eqref{KdV-BT-2}=& -6|\widehat{N-2}, \tau_{N}||\widehat{N-4}, N-2, N-1, N|
   -6|\widehat{N-4}, N-2, N, \tau_{N}||\widehat{N-1}| \\
~& +6|\widehat{N-2}, \tau_{N}||\widehat{N-3}, N-1, N+1|
  +6|\widehat{N-3},N+1, \tau_{N}||\widehat{N-1}| \\
~&-6|\widehat{N-3}, N-1,\tau_{N}||\widehat{N-2}, N+1|
   +6|\widehat{N-4}, N-2, N-1, \tau_{N}||\widehat{N-2}, N|,
\end{array}
\end{equation*}
which is zero in the light of Lemma \ref{Lemma-1}.

Finally, combining the Wronskian condition \eqref{Wrons-cond-2} and Lemma \ref{Lemma-3}
we can complete our proof for Theorem \eqref{Theor-1}.

\subsection{ Further discussion for the Wronskian condition}

In this subsection, we make a further discussion on the Wronskian
condition \eqref{KdV-BT-condition}.

First, noting that $f$ and $g$ in the bilinear BT \eqref{KdV-BT}
contribute solutions to the KdV equation \eqref{KdV} respectively through the transformation
\eqref{trans-f} and \eqref{trans-g},
one can conclude that for any $t$-dependent function $p(t)$,
$f$ and $p(t)\cdot f$ provide same solutions to the  KdV equation \eqref{KdV} through
\eqref{trans-f}, i.e., $u=2(\ln f)_{xx}=2[\ln (p(t)\cdot f)]_{xx}$. So do $g$ and $p(t)\cdot g$.

\begin{Lemma}
\label{Lemma-4}
~Suppose that matrices $g(\phi)$ and $f(\phi)$ are defined as \eqref{f-g}
with entry vector $\phi$;
\begin{equation}
\psi=P(t)\phi,
\label{phi-psi}
\end{equation}
where $P(t)=(P_{ij}(t))_{N\times N}$ is an non-singular matrix in
$\mathcal{G}_{N}[t]$. Then we have
\begin{equation}
g(\psi)=|P(t)|g(\phi),
~~f(\psi)=\frac{|P(t)|}{P_{NN}(t)}f(\phi).
\label{(aa)}
\end{equation}
\end{Lemma}
%
%

\noindent{\bf {Proof: }} Suppose that $P(t)=\Bigl(
    \begin{smallmatrix}
     \hat{P}(t)& ~{\bf 0}\\
     \rho(t) & ~P_{NN}(t) \\
    \end{smallmatrix}
   \Bigr). $
Then we have
$P^{-1}(t)
=\Bigl(
             \begin{smallmatrix}
              \hat{P}^{-1}(t) & 0 \\
              -\rho(t)\hat{P}^{-1}(t)P_{NN}^{-1}(t) & ~ P_{NN}^{-1}(t)\\
             \end{smallmatrix}
            \Bigr)
$
and $P^{-1}(t)\tau_{N}=(0, 0, \cdots, 0, P_{NN}^{-1}(t))^T$.
%
Then it is easy to check that
\begin{equation*}
\begin{array}{ll}
g(\psi)&=|P(t)\phi, P(t)\phi^{(1)}, \cdots, P(t)\phi^{(N-1)}|=|P(t)|g(\phi),\\
 f(\psi)& = |P(t)\phi, P(t)\phi^{(1)}, \cdots, P(t)\phi^{(N-2)}, P(t)P^{-1}(t)\tau_{N}|
 = |P(t)|P_{NN}^{-1}(t)f(\phi),
\end{array}
\end{equation*}
and thus we reach to \eqref{(aa)}.
\hfill$\Box$

Thus, under the condition of the above Lemma,
we conclude that $g(\phi)$ and $g(\psi)$
generate same solutions to the KdV equation \eqref{KdV}.
So do $f(\phi)$ and $f(\psi)$.

Next we turn to simplify the Wronskian condition \eqref{KdV-BT-condition}

For $B(t)$ in \eqref{Wr-Cd-b}, by virtue of Proposition \ref{Prop-2},
there exists a non-singular $t$-dependent matrix $H(t)$ in $\mathcal{G}_{N}[t]$
such that
\begin{equation}
H(t)_{t}=-H(t)B(t).
\end{equation}
If setting
\begin{equation}
\psi=H(t)\phi,
\end{equation}
\eqref{KdV-BT-condition} can be written as
\begin{equation}
\biggl\{\begin{array}{c}
 -\psi_{xx}=\widetilde{A}(t)\psi, \\
 \psi_{t}=-4\psi_{xxx},
\end{array}
\label{KdV-new-condition}
\end{equation}
where $\widetilde{A}(t)=H(t)A(t)H^{-1}(t)$ still belongs to $\mathcal{G}_{N}[t]$.
In the light of Lemma \ref{Lemma-4}, $\phi$ and $\psi$ lead to same solutions to the KdV equation.
That also means that we can drop $B(t)$ from \eqref{KdV-BT-condition},
and consequently the Wronskian condition \eqref{KdV-BT-condition}
is simplified to
\begin{equation}
\begin{array}{r}
 -\phi_{xx}=A\phi, \\
  \phi_{t}=-4\phi_{xxx},
\end{array}
\label{(2020)}
\end{equation}
where $A \in \mathcal{G}_{N}$ is arbitrary but independent of $t$ due to \eqref{KdV-AB}.

For further simplification, we can replace $A$ by its any similar matrix $\Gamma$
so long as the transform matrix $T$ is in $\mathcal{G}_{N}$, i.e.,
$\Gamma=T^{-1}AT$ where $T\in \mathcal{G}_{N}$.
We note that at this moment for any given matrix $A$ in
$\mathcal{G}_{N}$ we can not verify whether there exists a
non-singular matrix $B\in \mathcal{G}_{N}$ which makes $A$ into its
canonical form $\Gamma=B^{-1}AB$. However, by means of Proposition
\ref{Pro-4} we can employ a matrix $T \in \mathcal{G}_{N}$ to
transform $A$ into its quasi-canonical form $\Gamma$ and consider
the following Wronskian condition
\begin{equation}
\begin{array}{r}
- \psi_{xx}=\Gamma\psi, \\
  \psi_{t}=-4\psi_{xxx},
\end{array}
\label{final-cond}
\end{equation}
where $\psi=T\phi$,
\begin{equation}
\Gamma=\Biggl(
           \begin{array}{cc}
            \Gamma_{N-1} & {\bf 0} \\
             \gamma & a_{NN} \\
           \end{array}
         \Biggr )
\end{equation}
$\gamma=(\gamma_1,\gamma_2,\cdots,\gamma_{N-1})$
and $\Gamma_{N-1}$ is an $(N-1)$-order canonical matrix.



\section{Wronskian entries}

\subsection{General solutions to \eqref{final-cond}}

In this subsection, we discuss general solutions
to the condition equation \eqref{final-cond}.
We can separate \eqref{final-cond} into the following two sets:
\begin{equation}
\label{set-1}
{\rm Set~ I:}~~\biggl\{
\begin{array}{l}
 -\widetilde{\psi}_{xx}=\Gamma_{N-1} \widetilde{\psi},\\
 \widetilde{\psi}_{t}=-4\widetilde{\psi}_{xxx},
  \end{array}
\end{equation}
\begin{equation}
\label{set-2}
{\rm Set~ II:}~~\biggl\{
\begin{array}{l}
 -{\psi}_{N,xx}
 =a_{NN}{\psi}_{N}+\sum_{j=1}^{N-1}\gamma_{j}{\psi}_{j},\\
  {\psi}_{N,t}=-4{\psi}_{N,xxx},
  \end{array}
\end{equation}
where $\widetilde{\psi}=(\psi_1,\psi_2,\cdots,\psi_{N-1})^T$.
%
%

For Set I, since $\Gamma_{N-1}$ can be combinations of some diagonal matrices and Jordan blocks,
one can directly consider that $\Gamma_{N-1}$ is an $(N-1)$-order diagonal matrix with distinct
eigenvalues or an $(N-1)$-order Jordan block.
In Refs.\cite{Ma-You-KdV,Zhang-Wronskian},
Set I has been discussed in detail and explicit general solutions corresponding to different
$\Gamma_{N-1}$ are given.
Particularly, for $\Gamma_{N-1}$ being a Jordan block, the general solutions can be expressed in simple
forms by means of lower triangular Toeplitz matrices\cite{Zhang-Wronskian}.

For Set II, some formulae have been worked out for the solutions to the following
more general equations\cite{Ma-You-KdV},
\begin{equation}
\label{set-2-g}
\biggl\{
\begin{array}{l}
- {\psi}_{N,xx}
 =a_{NN}{\psi}_{N}+F,\\
  {\psi}_{N,t}=-4{\psi}_{N,xxx},
  \end{array}
\end{equation}
where the function $F=F(t,x)$ satisfies $F_t=-4 F_{xxx}$. By means of those formulae,
solutions to the Set II can easily be obtained.
Concretely, with $\widetilde{\psi}$ in hand, when $a_{NN}=0$, we have
\begin{equation}
- \psi_{N}=\int_0^x \int_0^{x'}F(t,
x^{\prime\prime})dx^{\prime\prime}dx'+c(t)x+d(t),
\end{equation}
where
\begin{equation}
c(t)=-4\int_0^tF_{xx}(t',0)dt'+c_0,~~~d(t)=-4\int_0^tF_{x}(t',0)dt'+d_{0},
\end{equation}
$F=\sum_{j=1}^{N-1}\gamma_{j}{\psi}_{j}$, both $c_0$ and $d_0$ are arbitrary real constants;
while when $a_{NN}$ begin a nonzero real or complex number we have
\begin{equation}
\begin{array}{rl}
-\psi_{N}=&\frac{1}{\sqrt{-a_{NN}}}\Bigl[(\sinh\sqrt{-a_{NN}}x)
\displaystyle\int_0^xF(x',t)(\cosh\sqrt{-a_{NN}}x')dx'\\
~&~~~~~~~~~~ -(\cosh\sqrt{-a_{NN}}x)\displaystyle\int_0^xF(x',t)(\sinh\sqrt{-a_{NN}}x')dx'\Bigr]\\
~&~~~~~~~~~~
+\frac{c(t)+d(t)}{2}(\sinh\sqrt{-a_{NN}}x)+\frac{c(t)-d(t)}{2}(\cosh\sqrt{-a_{NN}}x)
\end{array}
\end{equation}
where
\begin{equation*}
c(t)=e^{4a_{NN}\sqrt{-a_{NN}}t}\Bigl[c_0-4\int_0^te^{-4a_{NN}\sqrt{-a_{NN}}t'}(\frac{1}{\sqrt{-a_{NN}}}F_{xx}
+F_x+\sqrt{-a_{NN}}F)(0,t')dt'\Bigr],
\end{equation*}
\begin{equation*}
d(t)=e^{-4a_{NN}\sqrt{-a_{NN}}t}\Bigl[d_0+4\int_0^te^{4a_{NN}\sqrt{-a_{NN}}t'}(\frac{-1}{\sqrt{-a_{NN}}}F_{xx}
+F_x-\sqrt{-a_{NN}}F)(0,t')dt'\Bigr],
\end{equation*}
$F=\sum_{j=1}^{N-1}\gamma_{j}{\psi}_{j}$, both $c_0$ and $d_0$ are arbitrary complex constants.

General solutions to \eqref{final-cond} with some special $\gamma$ will be given in Appendix.

\subsection{Transformations of solutions}

If we consider $u=2 (\ln g)_{xx}$ as a new solution generated from the old one $u=2 (\ln f)_{xx}$,
obviously it is $\gamma$ and $a_{NN}$ to determine what the new solution we will get.
For example, if
\begin{equation*}
\Gamma={\rm Diag}(\lambda_{1}, \lambda_{2}, \cdots,\lambda_{N})
\end{equation*}
with $N$ distinct negative eigenvalues $\{\lambda_j=-k_j^2\}_{j=1}^{N}$, i.e.,
$(\gamma, a_{NN})=(0,0,\cdots,0,\lambda_{N})$,
we will get $N$ solitons from $(N-1)$ solitons.
If
\begin{equation*}
\Gamma=\left(
\begin{array}{ccccccc}
 \lambda_{1} & 0 & 0 & \cdots &0 & 0 & 0\\
 1 &  \lambda_{1}&0 & \cdots &0 & 0&0 \\
 \cdots & \cdots & \cdots & \cdots & \cdots & \cdots & \cdots\\
 0& 0 & 0 & \cdots &1 & \lambda_{1} & 0\\
 0& 0 & 0 & \cdots & 0 & 1 & \lambda_{1} \\
\end{array}
\right)_{N\times N},
\label{temp-1}
\end{equation*}
i.e.,
$(\gamma, a_{NN})=(0,0,\cdots,0,1,\lambda_{1})$, we will get an $N$-order Jordan block solution from
an $(N-1)$-order Jordan block solution. Particularly, if $\lambda_1=0$ in \eqref{temp-1} we get a transformation
between rational solutions.
If
\begin{equation*}
\Gamma=\left(
\begin{array}{ccccccc}
 \lambda_{1} & 0 & 0 & \cdots &0 & 0 & 0\\
 0 &  \lambda_{2}&0 & \cdots &0 & 0&0 \\
 \cdots & \cdots & \cdots & \cdots & \cdots & \cdots & \cdots\\
 0& 0 & 0 & \cdots &0 & \lambda_{N-1} & 0\\
 0& 0 & 0 & \cdots & 0 & 1 & 0 \\
\end{array}
\right)_{N\times N}
\end{equation*}
with $N-1$ distinct negative eigenvalues $\{\lambda_j=-k_j^2\}_{j=1}^{N-1}$,
i.e.,
$(\gamma, a_{NN})=(0,0,\cdots,0,1,0)$, we will get a mixed solution containing $(N-1)$ solitons and
one `rational part'.

Now we consider the transformations between complexitons\cite{Ma-complexiton} (or breathers\cite{breather-KdV}).
The $M$-complexiton solution corresponds to
$\Gamma$ is an even order canonical form and possesses $M$ complex conjugate couples of eigenvalues,
i.e.,
\begin{equation}
\Gamma={\rm Diag}(-k^2_1,-k^{*2}_1,-k^2_2,-k^{*2}_2,\cdots,-k^2_M,-k^{*2}_M),
\label{nor-matrix-Case6}
\end{equation}
where $\{-k^2_j\}$ are $M$ distinct complex numbers, and $*$ stands for complex conjugate.
To achieve the transformation from $(M-1)$ complexitons to $M$ complexitons,
we need two steps of BT, as shown below,
\begin{subequations}
\begin{eqnarray}
\label{T-1}
\Gamma={\rm Diag}(-k^2_1,-k^{*2}_1,-k^2_2,-k^{*2}_2,\cdots,-k^2_{M-1},-k^{*2}_{M-1})~~~~~~~~~~~\\
~~~~~~ \downarrow~ \hbox{\footnotesize $h=k^2_M$ in \eqref{KdV-BT}}\nonumber ~~~~~~~~~~~~~~~~~~~~~~~~~~\\
\Gamma={\rm Diag}(-k^2_1,-k^{*2}_1,-k^2_2,-k^{*2}_2,\cdots,-k^2_{M-1},-k^{*2}_{M-1},-k^2_M)~~~~~~\label{T-2}\\
~~~~~~ \downarrow~ \hbox{\footnotesize $h=k^{*2}_M$
in\eqref{KdV-BT}}\nonumber ~~~~~~~~~~~~~~~~~~~~~~~~~~\\
\Gamma={\rm
 Diag}(-k^2_1,-k^{*2}_1,-k^2_2,-k^{*2}_2,\cdots,-k^2_{M-1},-k^{*2}_{M-1},-k^2_M,-k^{*2}_M).
\label{T-3}
\end{eqnarray}
\end{subequations}
In other words, we first solve \eqref{KdV-BT} with $h=-k^2_M$,
$f=|\widehat{2M-3},\tau_{2M-1}|$ and $g=|\widehat{2M-2}|$ where the
Wronskian entries are provided by \eqref{final-cond} with $\Gamma$
defined by \eqref{T-2}. Then we solve \eqref{KdV-BT} with
$h=k^{*2}_M$, $f=|\widehat{2M-2},\tau_{2M}|$ and
$g=|\widehat{2M-1}|$ where the Wronskian entries are provided by
\eqref{final-cond} with $\Gamma$ defined by \eqref{T-3}. The
combination of these two BTs provides a transformation from $(M-1)$
complexitons expressed via $f=|\widehat{2M-3},\tau_{2M-1}|$ to $M$
complexitons expressed via $g=|\widehat{2M-1}|$.

In a similar way, we can achieve a transformation for high order complexitons (corresponding to
$\Gamma$ being a Jordan block defined as below). Suppose that
\begin{equation}
J_{2M}[\Lambda_{1}]=\left(
\begin{array}{ccccc}
\Lambda_{1} & 0 & \cdots & 0 & 0 \\
 I_{1} & \Lambda_{1} & \cdots & 0 & 0 \\
 \cdots & \cdots & \cdots & \cdots & \cdots \\
  0 & 0 & \cdots & I_{1} & \Lambda_{1}
  \end{array}
  \right)_{2M \times 2M},
\end{equation}
where
\begin{equation*}
I_{1}=\left(
       \begin{array}{cc}
        1 & 0 \\
        0 & 1 \\
       \end{array}
      \right),~~~
\Lambda_{1}=\left(
       \begin{array}{cc}
        -k^2_{1} & 0 \\
        0 & -k^{*2}_{1} \\
       \end{array}
      \right).
\end{equation*}
The Wronskian $g=|\widehat{2M-1}|$ where its entry vector $\psi$
solving \eqref{final-cond} with $\Gamma=J_{2M}[\Lambda_{1}]$
generates $M$-order complexitons\cite{Ma-complexiton}.
Then, a transformation between $(M-1)$-order complexitons and $M$-order complexitons
can be described as
\begin{equation*}
\begin{array}{c}
\Gamma=J_{2(M-1)}[\Lambda_{1}]~~~~~~~~~\\
~~~~~~~~~~~ \downarrow~ \hbox{\footnotesize $h=k^2_1$ in \eqref{KdV-BT}}\\
~~~~~~~~~~~ \Gamma=\left (\begin{array}{cc}
J_{2(M-1)}[\Lambda_{1}]& {\bf 0}\\
\gamma &  -k^2_1
\end{array}
\right )_{(2M-1) \times (2M-1)}~~~~~~~~~\\
~~~~~~~~~~~ \downarrow~ \hbox{\footnotesize $h=k^{*2}_1$ in \eqref{KdV-BT}}\\
\Gamma=J_{2M}[\Lambda_{1}],
\end{array}
\end{equation*}
where ${\bf 0}$ is a zero $(2M-2)$-order column vector and
$\gamma=(0,0,\cdots,0,1,0)$ is a $(2M-2)$-order row vector.

\vskip 12pt
\leftline{\bf Conclusions:} ~
We have discussed the Wronskian condition for the bilinear BT of the KdV equation.
The condition we obtained here is quite similar to the one for the bilinear KdV equation\cite{Zhang-Wronskian}.
We proved that this condition is reasonable by imposing suitable value to the parameter $h$
in the bilinear BT.
With further discussions the condition is finally the
differential equation set \eqref{final-cond} in which $\Gamma$ can be a
quasi-canonical matrix in $\mathcal{G}_N$.
General solutions for such a equation set have been given in Refs.\cite{Ma-You-KdV,Zhang-Wronskian}.
By choosing $\Gamma$ to be
some diagonals or Jordan blocks, the corresponding solution to \eqref{final-cond},
i.e., the Wronskian entry vector $\psi$, can provide solitons, negatons, positons, rational solution
and complexitons. Thus in the paper we have shown that the BT can
provide transformations between solitons and solitons, between negatons and negatons,
between positons and positons, between rational solution and rational solution,
and between complexitons and complexitons. It also admits transformations for mixed solutions.
Such discussions can be generalized to other soliton equations
with bilinear BT and Wronskian solutions.

\vskip 20 pt
\noindent{\bf Acknowledgments}~ This project is
supported by the National Natural Science Foundation of China
(10371070, 10671121), the Foundation of Shanghai Education Committee for Shanghai
Prospective Excellent Young Teachers, and Magnolia Grant of Shanghai Sciences and Technology Committee.

\begin{appendix}
\section{General solutions to \eqref{final-cond}}
\label{app-a}

Here for reference we list general solutions to \eqref{final-cond}
with $\Gamma$ being different canonical forms of an $N\times N$ constant matrix.
All these results we list here can be found in Refs.\cite{Zhang-Wronskian} and \cite{Ma-You-KdV}.

{\it Case 1}
\begin{equation}
\Gamma=D^{-}_{N}[\lambda_1,\lambda_2,\cdots,\lambda_N]
={\rm Diag}(-k^2_1,-k^2_2,\cdots,-k^2_N),
\label{nor-matrix-Case1}
\end{equation}
where  $\{ -k^2_j= \lambda_j \}$ are $N$ distinct negative numbers.
$\{ k_j\}$ are positive without loss of generality.
In this case,
\begin{equation}
\psi=\psi^{-}_{N}[\lambda_1,\lambda_2,\cdots,\lambda_N]
=(\psi^{-}_1,\psi^{-}_2,\cdots,\psi^{-}_N)^T,
\label{entry-I}
\end{equation}
in which
\begin{equation}
\psi^{-}_j=a^{+}_{j}\cosh {\xi_j} + a^{-}_{j}\sinh {\xi_j},
\label{entry-Case1}
\end{equation}
or
\begin{equation}
\psi^{-}_j=b^{+}_{j}e^{\xi_j} + b^{-}_{j}e^{-\xi_j},
\label{entry-Case1-1}
\end{equation}
where
\begin{equation}
\xi_j=k_j x -4k^3_j t +\xi^{(0)}_j,~~j=1,2,\cdots N,
\label{xi}
\end{equation}
$a^{\pm}_{j}$, $b^{\pm}_{j}$ and $e^{\xi^{(0)}_j}$  are all real constants.

{\it Case 2}
\begin{equation}
\Gamma=D^{+}_{N}[\lambda_1,\lambda_2,\cdots,\lambda_N]
={\rm Diag}(k^2_1,k^2_2,\cdots,k^2_N),
\label{nor-matrix-Case2}
\end{equation}
where  $\{ k^2_j= \lambda_j \}$ are $N$ distinct positive  numbers.
$\{ k_j\}$ are positive without loss of generality.
In this case,
\begin{equation}
\psi=\psi^{+}_{N}[\lambda_1,\lambda_2,\cdots,\lambda_N]
=(\psi^{+}_1,\psi^{+}_2,\cdots,\psi^{+}_N)^T,
\label{entry-II}
\end{equation}
in which
\begin{equation}
\psi^{+}_j=a^{+}_{j}\cos {\theta_j} + a^{-}_{j}\sin {\theta_j},
\label{entry-Case2}
\end{equation}
or
\begin{equation}
\psi^{+}_j=b^{+}_{j}e^{i\theta_j} + b^{-}_{j}e^{-i\theta_j},
\end{equation}
where
\begin{equation}
\theta_j=k_j x +4k^3_j t+ \theta^{(0)}_j ,~~j=1,2,\cdots N,
\label{theta}
\end{equation}
$a^{\pm}_{j}$, $b^{\pm}_{j}$ and $e^{\theta^{(0)}_j}$ are all real constants.

{\it Case 3}

\begin{equation}
\Gamma=J^{-}_{N}[\lambda_1]
=\left(\begin{array}{cccccc}
-k^2_1 & 0    & 0   & \cdots & 0   & 0 \\
1   & -k^2_1  & 0   & \cdots & 0   & 0 \\
\cdots &\cdots &\cdots &\cdots &\cdots &\cdots \\
0   & 0    & 0   & \cdots & 1   & -k^2_1
\end{array}\right)_N,
\label{nor-matrix-Case3}
\end{equation}
where $ -k^2_1=\lambda_1$ is a negative  number and $k_1$ also positive. In this case,
\begin{equation}
\psi=\psi^{J^{-}}_N[\lambda_1]=\mathcal{A}\mathcal{Q}^{+}_{0}+ \mathcal{B} \mathcal{Q}^{-}_{0},
~~\mathcal{A},\mathcal{B}\in \widetilde{G}_N,
\label{gen-sol-KdV}
\end{equation}
where
\begin{equation}
\mathcal{Q}^{\pm}_0=(\mathcal{Q}^{\pm}_{0,0},\mathcal{Q}^{\pm}_{0,1},\cdots,\mathcal{Q}^{\pm}_{0,N-1})^T,
~~
\mathcal{Q}^{\pm}_{0,j}=\frac{(-1)^j}{j!}\partial^{j}_{\lambda_1} b^{\pm}_1e^{\pm \xi_1},
\end{equation}
and $\widetilde{G}_N$ is a semigroup consisting of all of $N$-order lower triangular
Toeplitz matrices\cite{Zhang-Wronskian}.

We can also alternatively consider
\begin{equation}
\Gamma=\widehat{\Gamma}^{-}_N[k_1]=\left(\begin{array}{ccccccc}
-k^2_1 & 0    & 0   & \cdots & 0   & 0 & 0 \\
-2k_1   & -k^2_1  & 0   & \cdots & 0   & 0 & 0\\
-1   & -2k_1    & -k^2_1 & \cdots & 0   & 0 & 0\\
\cdots &\cdots &\cdots &\cdots &\cdots &\cdots &\cdots \\
0   & 0    & 0   & \cdots & -1 & -2k_1  & -k^2_1
\label{Gamma-k-KdV}
\end{array}\right)_N,
\end{equation}
which generates same solutions to the KdV equation as \eqref{nor-matrix-Case3} does.
In this case we have
\begin{equation}
\psi= \widehat{\psi}^{J^{-}}_N[k_1]=\mathcal{A}\widehat{\mathcal{Q}}^{+}_{0}+ \mathcal{B} \widehat{\mathcal{Q}}^{-}_{0},
~~~~\mathcal{A},~\mathcal{B} \in \widetilde{G}_N,
\label{gen-sol-KdV-k-3}
\end{equation}
where
\begin{equation}
\widehat{\mathcal{Q}}^{\pm}_0=(\widehat{\mathcal{Q}}^{\pm}_{0,0},\widehat{\mathcal{Q}}^{\pm}_{0,1},\cdots,
\widehat{\mathcal{Q}}^{\pm}_{0,N-1})^T,~~~
\widehat{\mathcal{Q}}^{\pm}_{0,j}=\frac{1}{j!}  \partial^{j}_{k_1}b^{\pm}_1 e^{\pm \xi_1}.
\label{vector-k-KdV-3}
\end{equation}

{\it Case 4}
\begin{equation}
\Gamma=J^{+}_{N}[\lambda_1]
=\left(\begin{array}{cccccc}
k^2_1 & 0    & 0   & \cdots & 0   & 0 \\
1   & k^2_1  & 0   & \cdots & 0   & 0 \\
\cdots &\cdots &\cdots &\cdots &\cdots &\cdots \\
0   & 0    & 0   & \cdots & 1   & k^2_1
\end{array}\right)_N,
\label{nor-matrix-Case4}
\end{equation}
where $ k^2_1=\lambda_1$ is a positive number and $k_1$ also positive.
In this case,
\begin{equation}
\psi=\psi^{J^{+}}_{\hbox{\tiny{\it N}}}[\lambda_1]
=\mathcal{A}\mathcal{P}^{+}_{0}+ \mathcal{B} \mathcal{P}^{-}_{0},
~~\mathcal{A},\mathcal{B}\in \widetilde{G}_N,
\label{gen-sol-KdV-4}
\end{equation}
with
\begin{equation}
\mathcal{P}^{\pm}_{0}=(\mathcal{P}^{\pm}_{0,0},\mathcal{P}^{\pm}_{0,1},\cdots,\mathcal{P}^{\pm}_{0,N-1})^T,
~~~\mathcal{P}^{\pm}_{0,j}=\frac{(-1)^j}{j!}\partial^{j}_{\lambda_1}
b^{\pm}_1 e^{\pm i \theta_1}.
\end{equation}
And for
\begin{equation}
\Gamma=\widehat{\Gamma}^{+}_N[k_1]=\left(\begin{array}{ccccccc}
k^2_1 & 0    & 0   & \cdots & 0   & 0 & 0 \\
2k_1   & k^2_1  & 0   & \cdots & 0   & 0 & 0\\
1   & 2k_1    & k^2_1 & \cdots & 0   & 0 & 0\\
\cdots &\cdots &\cdots &\cdots &\cdots &\cdots &\cdots \\
0   & 0    & 0   & \cdots & 1 & 2k_1  & k^2_1
\label{Gamma-k-KdV-4}
\end{array}\right)_N.
\end{equation}
we have
\begin{equation}
\psi=\widehat{\psi}^{J^{+}}_{\hbox{\tiny{\it N}}}[k_1]
=\mathcal{A}\widehat{\mathcal{P}}^{+}_{0}+ \mathcal{B} \widehat{\mathcal{P}}^{-}_{0},
~~~\mathcal{A}, \mathcal{B}\in \widetilde{G}_N,
\end{equation}
with
\begin{equation}
\widehat{\mathcal{P}}^{\pm}_{0}
=(\widehat{\mathcal{P}}^{\pm}_{0,0},\widehat{\mathcal{P}}^{\pm}_{0,1},\cdots,\widehat{\mathcal{P}}^{\pm}_{0,N-1})^T,
~~~\widehat{\mathcal{P}}^{\pm}_{0,j}=\frac{1}{j!}\partial^{j}_{k_1}
b^{\pm}_1 e^{\pm i \theta_1}.
\end{equation}

{\it Case 5} ~
\begin{equation}
\Gamma=J^{0}_{N}
=\left(\begin{array}{cccccc}
0 & 0    & 0   & \cdots & 0   & 0 \\
1   & 0  & 0   & \cdots & 0   & 0 \\
0   & 1  & 0   & \cdots & 0   & 0 \\
\cdots &\cdots &\cdots &\cdots &\cdots &\cdots \\
0   & 0    & 0   & \cdots & 1   & 0
\end{array}\right)_N.
\label{nor-matrix-Case5}
\end{equation}
In this case,
\begin{equation}
\psi=\psi^{0}_N=\mathcal{A}\mathcal{R}^{+}_{0}+ \mathcal{B} \mathcal{R}^{-}_{0},
~~\mathcal{A},\mathcal{B}\in \widetilde{G}_N,
\label{gen-sol-KdV-5}
\end{equation}
where
\begin{equation}
\mathcal{R}^{\pm}_0=(\mathcal{R}^{\pm}_{0,0},\mathcal{R}^{\pm}_{0,1},\cdots,\mathcal{R}^{\pm}_{0,N-1})^T
\end{equation}
with
\begin{equation}
R^{+}_{0,j}=\frac{(-1)^j}{(2j)!}\Bigl [\frac{\partial^{2j}}{{\partial k_1}^{2j}}\cosh \xi_1\Bigr ]_{k_1=0},~~~
R^{-}_{0,j}=\frac{(-1)^j}{(2j+1)!}\Bigl [\frac{\partial^{2j+1}}{{\partial k_1}^{2j+1}}\sinh \xi_1\Bigr ]_{k_1=0}.
\end{equation}

{\it Case 6} ~
\begin{equation}
\Gamma=D^{\hbox{\it\tiny C}}_{\hbox{\tiny 2\it M}}[\lambda_1,\lambda_2,\cdots,\lambda_M]
={\rm Diag}(-k^2_1,-k^{*2}_1,-k^2_2,-k^{*2}_2,\cdots,-k^2_M,-k^{*2}_M),
\label{nor-matrix-Case6}
\end{equation}
where $\{-k^2_j=\lambda_j\}$ are $M$ distinct complex numbers, and $*$ means complex conjugate.
In this case,
\begin{equation}
\psi=\psi^{\hbox{\it\tiny C}}_{\hbox{\tiny 2\it M}}[\lambda_1,\lambda_2,\cdots,\lambda_{M}]
=(\psi^{\hbox{\it\tiny C}}_1, \psi^{\hbox{\it\tiny C}*}_1,
\psi^{{\hbox{\it\tiny C}}}_2, \psi^{{\hbox{\it\tiny C}}*}_2,\cdots,
\psi^{\hbox{\it\tiny C}}_{\hbox{\it\tiny M}}, \psi^{\hbox{\it\tiny C}*}_{\hbox{\it\tiny M}})^T,
\label{entry-Case6-0}
\end{equation}
where
\begin{equation}
\psi^{\hbox{\it\tiny C}}_j=a^{+}_{j}\cosh {\xi_j} + a^{-}_{j}\sinh {\xi_j},
\end{equation}
or
\begin{equation}
\psi^{\hbox{\it\tiny C}}_j=b^{+}_{j}e^{\xi_j} + b^{-}_{j}e^{-\xi_j},
\label{entry-Case6-1}
\end{equation}
with
\begin{equation}
\xi_j=k_j x -4k^3_j t +\xi^{(0)}_j,~~j=1,2,\cdots M,
\label{xi-6}
\end{equation}
$a^{\pm}_{j}$, $b^{\pm}_{j}$ and $e^{\xi^{(0)}_j}$  are all complex constants.

{\it Case 7} ~
\begin{equation}
\Gamma= J^{\hbox{\it\tiny C}}_{2M}[\lambda_1]=\biggl(\begin{matrix}\Lambda & 0\\
                                 0  & \Lambda^* \end{matrix}\biggr),
~~\Lambda
=\left(\begin{array}{ccccc}
-k^2_1 & 0    &  \cdots & 0   & 0 \\
1   & -k^2_1  &  \cdots & 0   & 0 \\
\cdots &\cdots &\cdots &\cdots  &\cdots \\
0   & 0    &  \cdots & 1   & -k^2_1
\end{array}\right)_{2M},
\label{(2.56)}
\end{equation}
where  $-k^2_1=\lambda_1$.
In this case,
\begin{equation}
\psi
=\Biggl(\begin{array}{cc}
\mathcal{A}& 0\\
0& \mathcal{A}^*
\end{array}\Biggr)\Biggl(
\begin{array}{c}
\mathcal{Q}^{+}_{0}\\
{\mathcal{Q}^{+}_{0}}^*
\end{array}\Biggr)
+\Biggl(\begin{array}{cc}
\mathcal{B}& 0\\
0& \mathcal{B}^*
\end{array}\Biggr)\Biggl(
\begin{array}{c}
\mathcal{Q}^{-}_{0}\\
{\mathcal{Q}^{-}_{0}}^*
\end{array} \Biggr),
~~ \mathcal{A}, \mathcal{B} \in \widetilde{G}_M,
\label{gen-sol-KdV-c}
\end{equation}
where
\begin{equation}
\mathcal{Q}^{\pm}_0=(\mathcal{Q}^{\pm}_{0,0},\mathcal{Q}^{\pm}_{0,1},\cdots,\mathcal{Q}^{\pm}_{0,M-1})^T,
~~
\mathcal{Q}^{\pm}_{0,j}=\frac{(-1)^j}{j!}\partial^{j}_{\lambda_1} b^{\pm}_1e^{\pm \xi_1}.
\end{equation}

If we alternatively take
\begin{equation}
\Gamma=\check{J}^{\hbox{\it\tiny C}}_{2M}[K_1]
=\left(\begin{array}{ccccccc}
-K^2_1 & 0   & 0  &  \cdots & 0   & 0 & 0 \\
-2K_1   & -K^2_1  & 0& \cdots & 0   & 0 & 0 \\
-I_1   & -2K_1  & -K^2_1 & \cdots & 0   & 0 & 0 \\
\cdots &\cdots &\cdots & \cdots &\cdots  &\cdots & \cdots\\
0   & 0  & 0  &  \cdots & -I_1 & -2K_1   & -K^2_1
\end{array}\right)_{2M},
\label{J-KdV-7-3-a}
\end{equation}
where
\begin{equation}
K_1=\biggl(\begin{matrix}k_1 & 0\\
                                 0  & k_1^* \end{matrix}\biggr),
\label{J-KdV-7-3-b}
\end{equation}
and $-K^2_1=\Lambda_1$, we have
\begin{equation}
\psi=\check{\psi}^{J_C}_{\hbox{\tiny 2\it M}}[k_1]
=\mathcal{A}^B\check{\mathcal{Q}}^{+}_{0}+
\mathcal{B}^B \check{\mathcal{Q}}^{-}_{0},
\label{gen-sol-KdV-c-7-3}
\end{equation}
with
\begin{equation}
\check{\mathcal{Q}}^{\pm}_0=(\check{\mathcal{Q}}^{\pm}_{0,0},\check{\mathcal{Q}}^{\pm *}_{0,0},
\check{\mathcal{Q}}^{\pm}_{0,1},\check{\mathcal{Q}}^{\pm *}_{0,1},\cdots,
\check{\mathcal{Q}}^{\pm}_{0,M-1},\check{\mathcal{Q}}^{\pm *}_{0,M-1})^T,
~~
\check{\mathcal{Q}}^{\pm}_{0,j}=\frac{1}{j!}\partial^{j}_{k_1} b^{\pm}_1e^{\pm \xi_1},
\label{Q-check}
\end{equation}
where $\mathcal{A}^B$ and $\mathcal{B}^B$ are $2M$-order block lower triangular
Toeplitz matrices with  entries $A_j$ and $B_j$ defined as
\begin{equation}
A_j = \biggl(\begin{matrix} a_{j1} & 0\\
                                 0  & a^*_{j1} \end{matrix}\biggr),~~~
B_j = \biggl(\begin{matrix} b_{j1} & 0\\
                                 0  & b^*_{j1} \end{matrix}\biggr).
\end{equation}

\end{appendix}

\end{document}